# Amoeboid movement utilizes the shape coupled bifurcation of an active droplet to boost ballistic motion

H. Ebata[1], Y. Nishigami[2,3], H. Fujiwara[4], S. Kidoaki[5], M. Ichikawa[4, *]

[1] Department of Physics, Kyushu University, 744 Motooka, Nishi-ku, Fukuoka 819-0395, Japan

[2] Global Station for Soft Matter, Global Institution for Collaborative Research and Education, Hokkaido University, Sapporo, 001-0021, Japan

[3] Research Institute for Electronic Science, Hokkaido University, N20 W10, Kita-ku, Sapporo, 001-0020, Japan

[4] Department of Physics, Graduate School of Science, Kyoto University

[5] Institute for Materials Chemistry and Engineering, Kyushu University, 744 Motooka, Nishi-ku, Fukuoka 819-0395, Japan

Corresponding author: * M. Ichikawa

Email: ichi@scphys.kyoto-u.ac.jp

**Abstract**
One of the essential functions of living organisms is spontaneous migration through the deformation of their body, such as crawling, swimming, and walking. Depending on the size of the object, the efficient migratory mode should be altered because the contribution from the inertial and frictional forces acting on the object switches. Although the self-propelling motion characterizing active matter has been extensively studied, it is still elusive how a living cell utilizes the mode switching of the self-propulsion. Here, we studied the migration dynamics of amoeboid movement of free-living amoeba, *Amoeba proteus*, for starved and vegetative phases, as typified by dynamic and stationary states, respectively. Fourier-mode analysis on the cell shape and migration velocity extracted two characteristic migration modes, which makes a coexistence of amoeboid-swimmer like random motion and the active-droplet like ballistic motion. While the amoeboid-swimmer mode governs random motion, the active-droplet mode performs non-negligible contribution on the migration strength. By employing the symmetry argument of the active-droplet, we discover the supercritical pitchfork bifurcation of the migration velocity due to the symmetry breaking of the cell shape represents the switching manner from the motionless state to the random and the ballistic motions. Our results suggest that sub-mm sized *A. proteus* utilizes both shape oscillatory migration of deformed-swimmer driven by surface wave and convection based mass transfer, called blebbing, as like as cm-sized active droplet to optimize the movement efficiency.

**Significance Statement**
Self-propulsion, which is ubiquitous among living organisms and non-living active materials, requires spatial and temporal symmetry-breaking. Fourier-mode analysis on the cell shape of sub-mm sized *Amoeba proteus* extracts the coexistence of amoeboid-swimmer mode and active-droplet modes, and moreover represents supercritical pitchfork bifurcation from stationary state to dynamic state due to the symmetry breaking of the cell shape. While the amoeboid-swimmer mode regulates the shape oscillation based random migration, the active-droplet modes contribute ballistic motion. The present result indicates a basic symmetry law of self-propulsion underlying living cells and non-living active materials.

**Introduction**
Self-propulsions such as swimming, crawling and walking are characteristic features and crucial function of living organisms and non-living active materials (1). Depending on the surrounding environment and spatial scale of the objects typically ranging from $10^{-7}$ to $10^0$ m, optimal physical processes involved in the self-propulsion are qualitatively different in presence of frictional force on the substrate (2), the ratio between inertial force and viscous drag force (3), and the influence of thermal and athermal noise (4). While the mechanism of self-propulsion has been studied for each situation, it is still unclear whether the universal mathematical and physical architecture underlies the self-propulsion in both living and non-living matters.

To drive the self-propulsion of an object, spatial and temporal symmetry-breaking must arise on surface property (5, 6), chemical concentration field (7), and shape of the object (8, 9). In any case, dynamics of the self-propulsion are crucially regulated by what kind of the symmetry breaks; front-rear and rotational asymmetry cause the straight and rotational motions (5, 10-13), respectively. In general, bifurcation coincides with symmetry breaking, and universal feature appears around the bifurcation point. Thus, regardless of living or non-living, universal laws of self-propulsion should emerge when a mode of the motion changes beside the spatial symmetry breaking. Although bifurcation of self-propulsion has been investigated in non-living systems (14-17), it has been difficult to quantitatively study that of living organisms because of the complex response to the control parameters.

*Amoeba proteus* is one of the candidates to investigate the qualitative and quantitative change of the motion because we can control the mode of the migration depending on the duration time after the feeding (18-21). Just after feeding, *A. proteus* becomes stationary with circular shape, while they show the persistent random motion (22) in the starved phase, called amoeboid movement. In the movement of *A. proteus*, contraction of actomyosin cortex drives a protoplasmic flow to deform the cell body and to propel the cell (23, 24). Such persistent random motion has been observed for various types of cells. For mesenchymal cells, migration direction of the persistent random motion synchronize long axis direction of the deforming cell shape (25), which should correlate to the dynamics of internal actin flow (26). On the other hand, random and fluctuating motion comes from the extension and contraction of the cell pseudopodia (25). Although the previous studies indicate that the dynamics of the cell shape regulate the cell movement (8, 27-33), it has not been revealed what kind of the symmetry breaking of the cell shape is essential to the bifurcation of the motion.

The present study aimed to reveal the relation between mode bifurcation of the single cell migration and the dynamics of cell shape. Through a Fourier-mode analysis of the shape (25), we investigated quantitative relationship between velocity and shape of *A. proteus*. By using the velocity equation based on the model of soft active objects (25, 34-36), we showed that the ballistic migration underwent supercritical bifurcation from stationary state to the ballistic motion state due to the symmetry breaking of the cell shape, while deformation swimmer motion (25, 37) bears the persistent random motion of the migration. The proposed velocity equation is generally applicable to the self-propelled motion of living and nonliving soft materials from micro to macro, where coupled system of deformation swimming and interface-convection-driven swimming such as trembling droplet (35), propelling pattern in suspension (38), amoeboid swimmer (37), and crawling mesenchymal cells (25).

**Results**

Figure 1(a) shows typical results of translational motions and shape deformations of *A. proteus* at day 0 (left) and at day 1 (right). The number of days is the spending period from feeding. To regulate deformation activity relating to the cell shape and the translational motion, we changed the duration time to start the measurement after feeding. From the left side to the right side in Fig. 1 (b), trajectories of migration motions are collected for the cells of 0, 1, and 3 days after feeding, respectively. Just after the feeding (0 day under starvation condition), *A. proteus* was almost stationary and had circular shape (see Supplementary movie). As the time went on (1 to 3 days), *A. proteus* started to exhibit persistent random motion that consisted of straight, ballistic and random, fluctuating turns (see Supplementary movie). Note that the duration time after feeding did not completely determine the onset of the migration; a few cells showed stationary at 3 days under starvation condition. Since the feeds surrounded had been removed experimentally before the 0-day observation, and since the observation light was deep red, globally biased motions such as chemotaxis and phototaxis were not observed (39, 40).

Before studying the bifurcation of the motion, we first separately characterize two distinct movements, persistent random and ballistic movements, to reveal what component was prominent, and then link both sides accompanied by all other data. Thus, for the characterization of the cell movement, we focused on the relation between cell velocity and dynamics of the cell shape by using the Fourier modes analysis (25). The complex Fourier coefficient $C_n$ of the shape was calculated from the cell-shape $R$, where $R(\theta)$ denoted the distance from the cell centroid to the rim as a function of the angle $\theta$ measured from $x$ axis (Fig. 2 (a)).

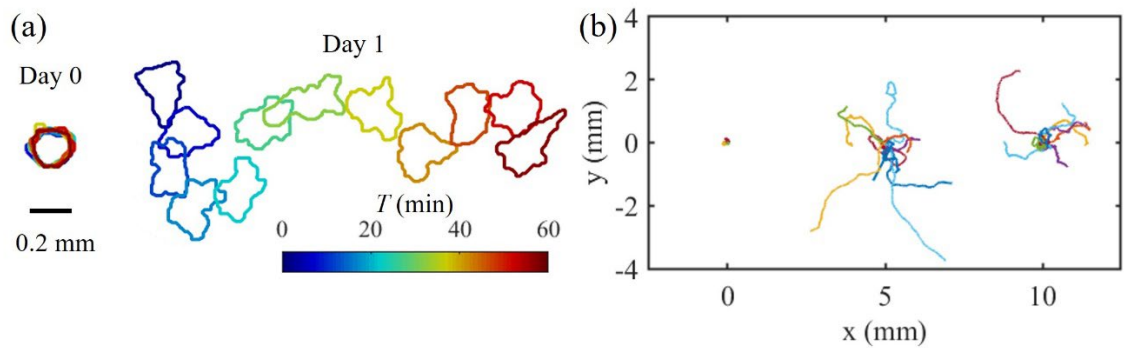

**Figure 1.** (a) Representative time series of a migration of amoeba for 0 day (left) and 1 day (right) after feeding. Each frame is taken every 300 second. (b) Trajectories of the cell centroids. From left to right sides, we showed trajectories of 0, 1, and 3 days after feeding, respectively. The initial point was set to be identical.

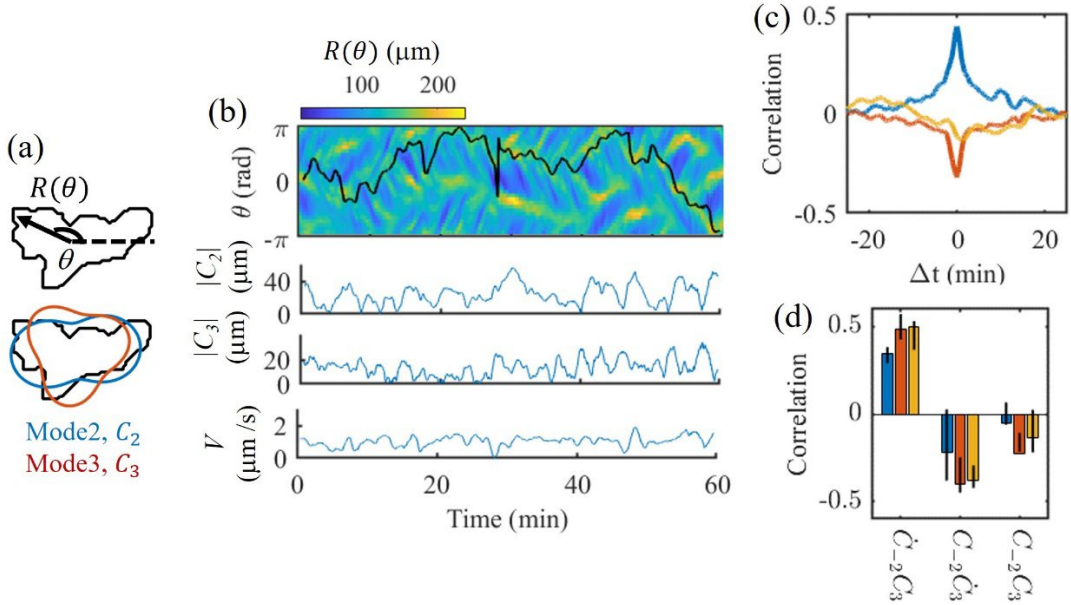

**Figure 2.** (a) Top: periphery of the cell detected by image analysis. Cell shape $R$ was a function of angle $\theta$. Bottom: Elongation $C_2$ (blue line) and triangular deformation $C_3$ (red line) were measured through Fourier mode analysis of the periphery. (b) From upper to lower figures, figures show typical spatiotemporal plot of shape $R(\theta, t)$, time courses of mode 2 $|C_2|$, mode 3 $|C_3|$, and cell speed $V$, respectively. Color bar represents magnitude of $R$. Black line in spatiotemporal plot indicates direction of the migration. (c) Cross correlation between cell velocity and non-linear terms of $C_2$ and $C_3$. Blue: $\dot{C}_{-2}C_3$. Red: $C_{-2}\dot{C}_3$. Yellow: $C_{-2}C_3$. (d) Cross correlation at $\Delta t = 0$. Bar indicates a median value. Error bar connects the 0.25 and 0.75 quantiles. Color indicates the duration time after feeding. Blue: 0 day. Red: 1 day. Yellow: 3 days.

$$R(\theta) = R_0 + \sum_{n=2}^{m} \left( C_n(t)e^{in\theta} + C_{-n}(t)e^{-in\theta} \right), \quad (1)$$

where $R_0$ is the mean radius and $m$ is the number of data points. Then, time evolution of $C_n$ quantitatively characterized dynamics of the cell shape.

First, we characterized the shape fluctuation of the migration with the persistent random mode. From upper to lower figures, Fig. 2 (b) shows typical spatiotemporal plots of the shape $R(\theta, t)$, time courses of mode 2 $|C_2|$, mode 3 $|C_3|$, and cell speed $V$, respectively. The figures indicate the characteristics that *A. proteus* randomly extended and contracted the pseudopodia (see Supplementary movie). As a result, both Fourier modes of the shape and cell speed fluctuated. When the fluctuation of cell shape regulated the migration of the cell, time derivative of $C_n$ should correlate to the cell velocity (25). From the argument of spatial symmetry, we provided the possible relations between velocity and $C_n$ with lowest modes and lowest nonlinearity (25, 36).

$$v_1 = v_{fluc} + \beta_1 \dot{C}_{-2}C_3 - \beta_2 C_{-2}\dot{C}_3 + \beta_c C_{-2}C_3, \quad (2)$$

where $v_1 = v_x + i v_y$ is complex velocity of the cell, and $C_{-n}$ is complex conjugate of $C_n$. The equation is a combination of deformation-swimming mode that is called amoeboid-swimmer (25, 37) and keratocyte-like migration mode $\beta_c C_{-2} C_3$ (36). To confirm the validity of Eq. (2), we calculated the cross-correlation between cell velocity and the nonlinear terms of $C_2$ and $C_3$ (Fig. 2 (c)). Cell velocity had moderate correlation with $\dot{C}_{-2} C_3$ and $C_{-2} \dot{C}_3$ at $\Delta t = 0$ (correlation coefficient ~ 0.5), while $C_{-2} C_3$ was less correlated (correlation coefficient ~ 0.2). This tendency held regardless of the duration time after feeding in the ensemble data analysis (Fig. 2 (d)). However, when we calculated the coefficient of the determination $R^2$ of Eq. (2), 10% of the cells satisfies $R^2 > 0.4$. Thus, contribution of the shape fluctuation to the migration Eq. (2) was not sufficient to reproduce the locomotion of *A. proteus*. Since the first and second terms of Eq. (2) corresponds to the efficiency of the fluctuating motion by the extension and contraction of the cell body, the results suggest that straight and ballistic movements also play an important role in the migration of *A. proteus*.

In addition to the movement above mentioned, a small number of *A. proteus* also showed ballistic movement with almost constant shape especially under 1 and 3 day starvation conditions (see Supplementary movie). From upper to lower figures, Fig. 3 (a) shows typical spatiotemporal dynamics between the shape and migration speed $V$ for *A. proteus* with the ballistic movement. In contrast to the case of the fluctuating cell shape (Fig. 2 (b)), amplitude of mode 2 $|C_2|$ was almost constant, and amplitude of mode 3 $|C_3|$ was small. Since yellow colored ridge lines in the spatiotemporal plot of $R(\theta, t)$ correspond the position of both edges of the elongated shape, Fig. 3 (a) indicates that the cell kept a highly elongated shape with small fluctuation. The amoeboid-swimmer like velocity equation Eq. (2) cannot represent alone the essential feature of the ballistic motion with constant shape, because $\dot{C}_2$ and $\dot{C}_3$ almost vanished.

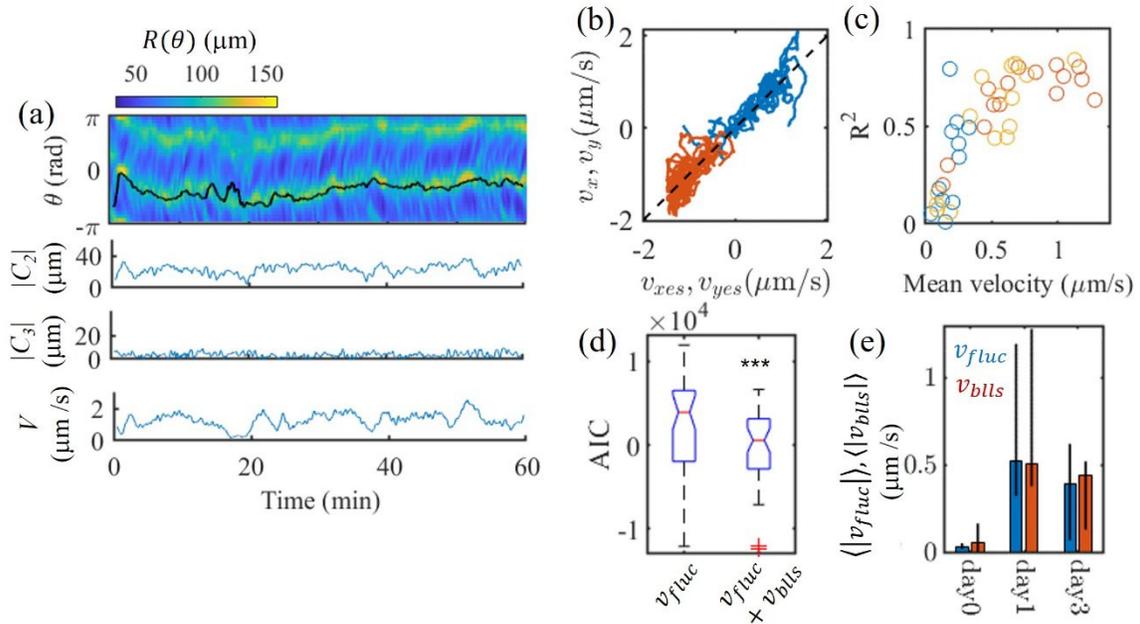

**Figure 3.** (a) From upper to lower figures, we show typical spatiotemporal plot of shape $R(\theta, t)$, time courses of mode 2 $|C_2|$, mode 3 $|C_3|$, and cell speed $V$ with ballistic movement, respectively. Color bar represents magnitude of $R$. Black line in the spatiotemporal plot indicates direction of the migration. (b) Correlation between velocity $(v_{xes}, v_{yes})$ estimated from Eq. (4) and actual cell velocity. The data was identical to Fig. 3 (a). Blue: $x$ component of velocity. Red: $y$ component of velocity. (c) Coefficient of determination $R^2$ of Eq. (4) as a function of mean cell speed. Each symbol represents the data calculated from a time course of a cell. Color indicates the duration time after feeding. Blue: 0 day. Red: 1 day. Yellow: 3 days. (d) Boxplot of Akaike information criteria (AIC) for $v_{fluc}$ and $v_{fluc} + v_{blls}$. The paired t-test was used to calculate the P value. ***: p < 0.001. (e) Contribution of $v_{fluc}$ (blue) and $v_{blls}$ (red) in Eq. (4) on fitted velocity. Bars indicate the median values of $\langle |v_{fluc}| \rangle$ and $\langle |v_{blls}| \rangle$. Error bar connects the 0.25 and 0.75 quantiles.

To explain the ballistic mode of *A. proteus*, another velocity equation is needed. The migration of *A. proteus* had two characteristic features. First, migratory cells had more elongated shape than the stationary cells had (Fig. 1 (a)). Second, as shown in Fig. 3 (a), the direction of the migration velocity (black solid line) highly correlated to the direction of elongation (yellow colored ridge lines). These features imply strong coupling between velocity $v_1$ and elongation $C_2$. In the model of soft active particles that contains coupling term of $v_1$ and $C_2$, large elongation of the body causes the bifurcation from stationary state to ballistic movement (35). Thus, we examined whether the elongation-induced movement could be applied for the ballistic movement and for all about the movement of *A. proteus*. The velocity equation describing elongation-induced movement (35) is

$$v_{blls} = \alpha_1 v_{-1} C_2 - \gamma v_1 |v_1|^2. \tag{3}$$

Then, we estimate the cell velocity by the combination of fluctuating and ballistic movement in maximum likelihood methodology,

$$v_1 = v_{fluc} + v_{blls}. \tag{4}$$

Figure 3 (b) shows an example of the correlation between fitted velocity by using Eq. (4) and actual cell velocity. In the fitting procedure, we numerically search the least square error by changing the coefficients in the model. Both $x$ and $y$ component of fitted velocity had high correlation with that of actual velocity (correlation > 0.8), indicating that Eq. (4) can more precisely predict the cell velocity. Then, we calculated the coefficient of determination $R^2$ of the migration model Eq. (4) for each cell. As shown in Fig. 3 (c), regardless of the duration time after feeding, $R^2$ increased to about 0.8 as the averaged cell speed increased, and about 70 % of the cells satisfies $R^2 > 0.4$.

To quantitatively compare capability of the model equations, we calculated the Akaike information criteria (AIC) (41) of Eqs. (2) and (4), where smaller AIC gives a better explanation. AIC of Eq. (4) was significantly smaller than AIC of Eq. (2) (Fig. 3 (d)). The above results denote that the accuracy of the velocity equation increases when we include the model of the ballistic movement, which suggests coexistence of driving forces caused by the amoeboid-swimmer type and by the active-droplet type. To evaluate the contribution from two driving terms, we calculated the averaged value of $v_{fluc}$ and $v_{blls}$ in Eq. (4) by using the fitted coefficients $\beta_i$, $\alpha_1$, and $\gamma$ (Fig. 3 (e)). Then, we found that the active droplet term Eq. (3) and amoeboid-swimmer term Eq. (2) had close values. Therefore, both fluctuating and ballistic terms almost equally contributed to the velocity $v_1$.

The model of active droplet exhibiting ballistic movement Eq. (3) is reported to show the supercritical pitchfork bifurcation when the magnitude of elongation exceeds the critical value (35). Thus, we examined whether the deformation-induced bifurcation of the movement was responsible for the motion transition phenomenologically observed in the amoebous motion (Fig. 1). Bifurcation of the stable solution of Eq. (3) is

$$|v_1|^2 = 0, \qquad |C_2| < A \tag{5}$$
$$|v_1|^2 = (\alpha_1 |C_2| \cos 2\psi_2 - 1)/\gamma, \qquad |C_2| \geq A \tag{6}$$
$$A = 1/\alpha_1 \cos 2\psi_2, \tag{7}$$

where phase difference $\psi_2$ is defined as $\psi_2 = \phi_2 - \phi_v$ with $\phi_v = \arg(v_1)$ and $\phi_2 = \arg(C_2)/2$. By using the estimated coefficients in Eq. (4) through the fitting, we tested relation between cell speed and elongation (Fig. 4). Here, to concern only with the long-term behavior of the cell migration, we calculated the cell speed with large time interval $\Delta t = 900$ s. Then, we averaged the speed and elongation over a time course.

Figure 4 (a) shows the experimental data plotted with axes of the cell speed and of the degree of elongation as expected from Eqs. (5) and (6). The data exhibit supercritical pitchfork bifurcation and are aligned on the lines calculated by the theoretical model with parameters obtained in Fig. 3. These results show that time averaged velocity $\langle v_1(t) \rangle_t$ obeyed active droplet equation Eq. (3), while amoeboid-swimmer term Eq. (2) was needed to determine instantaneous velocity $v_1(t)$ (Fig. 3 (e)).

While length of the duration time after feeding did not preciously explain the start time of the movement for each cell, the parameters such as $\alpha_1$, $|C_2|$, and $\cos 2\psi_2$ which gradually changed on the duration explained the bifurcation behaviors quantitatively through Eqs. (5) to (6) (Figs. 4 (b) – (d)). Increase of $\alpha_1$ indicated that the coupling between velocity and elongation was enhanced as the duration time went on. The coupling coefficient $\alpha_1$ should cause the alignment of directions of velocity and elongation (25) as observed in Figs. 2 (a), 3 (a), and 4 (d), which also made motionless state unstable (35). This fact strongly supports the consistency and applicability of the present model: When the elongation of the cell shape exceeds the critical value, symmetry breaking of an internal factor arises to cause the supercritical pitchfork bifurcation of the migration speed.

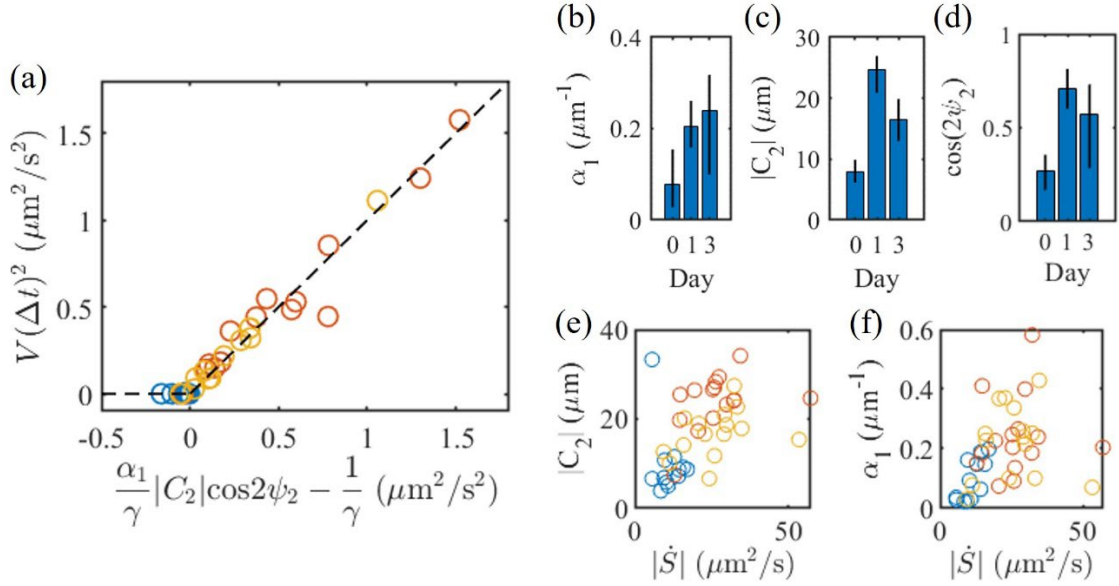

**Figure 4.** (a) Relation between steady part of cell speed $V(\Delta t)$ and the magnitude of elongation of a cell $|C_2|$. $\psi_2$ denotes phase difference between cell velocity $v_1$ and elongation $C_2$. Black dashed line represents theoretical prediction from Eq. (5) and (6). The coefficients $\alpha_1$ and $\gamma$ were estimated from the fitting of Eq. (4). $V$, $|C_2|$ and $\cos 2\psi_2$ were averaged over a time course of a cell motion. Each data point denotes the data of a cell. $\Delta t = 900$ s. (b - d) Dependence of the parameters on the duration time after feeding. (b) $\alpha_1$. (c) $|C_2|$. (d) $\cos 2\psi_2$. Bar indicates a median value. Error bar connects the 0.25 and 0.75 quantiles. (e) Correlation between time variation of cell Area $|\dot{S}|$ and $|C_2|$. Correlation coefficient is 0.49. (f) Correlation between $|\dot{S}|$ and $\alpha_1$. Correlation coefficient is 0.38. (a, e, f) Color indicates the duration time after feeding. Blue: 0 day. Red: 1 day. Yellow: 3 days.

**Discussion**

Although physical meaning of the shape and its time derivative terms has not been explained theoretically, we suggest that one of possible internal factors observed beside movement of *A. proteus* is an internal flow of the protoplasm (42, 43). Interestingly, the non-living cm-sized swimmer is reported to obey Eq. (3); the swimming droplet driven by the surface wave undergoes the same bifurcation of the elongation-induced movement (35). Since stability of convectional flow in the droplet is coupled to the magnitude of the elongation of the body through the radiation pressure from the surface wave, symmetry breaking of the surface wave leads to asymmetric convection that drives ballistic motion of the droplet. In spherical droplet swimmers, asymmetry of the surface tension plays the same role instead of shape anisotropy to cause a dipole flow (44). On the other hand, in the migrating cells, interplay between actin flow and polarity cue is reported to regulate the speed and persistence of the cell motion (26). Since the cell polarity is coupled with the elongation of the cell, actin flow should interact with elongation of the body through the polarity cue. The corresponding fluid convection is also reported for *A. proteus* that protoplasmic flow is generated not only inside pseudopodia but in the whole body with longitudinal direction driven by contraction of posterior actomyosin (42, 43, 45). Assuming that the speed of the cell or droplet is proportional to the speed of the internal flows in respect to mass transfer (26, 35), the term $\alpha_1 C_2 v_{-1}$ in Eq. (3) represents the coupling between the internal flow and elongation. In this scheme, $\alpha_1$ should represent the magnitude of coupling among the flow, polarity cue, and elongation, and hidden parameters such as size and mass, etc., affect through them. As shown in Figs. 4 (e) and (f), we found that the time variation of cell area $|\dot{S}|$ had positive correlation with $|C_2|$ and $\alpha_1$. Assuming that the cell volume was constant, change of $S$ corresponded to the variation of cell height, which could cause the flow in the cell. It suggested that 'pumping motion' of the cell enforced the elongation of blebbing accompanied with violation of front-rear symmetry.

The proposed migration model Eq. (4) had general formulation of the velocity equation that includes models of non-living and living active soft matters with both macro and micro scales, such as convection-driven swimming droplet (35), propelling pattern in suspension (38), amoeboid swimmer (37), blebbing actomyosin gel (43), and crawling mesenchymal cells on the substrate (25). In those experiments, the self-propulsion and the bifurcation of the motion can be reproduced by the non-linear equations of the velocity and deformations. It implies that regardless of living or non-living, the difference among the deformation-driven self-propulsions only arises on the value of the coefficients. Consequently, the results suggest that when we focus on the deformation-induced movement, both living and non-living active matters undergo the same cascade of mode bifurcation accompanied with the symmetry breaking of the shape.

## Materials and Methods

### Cell culture

*Amoeba proteus* was cultured in KCM medium (9.39 μM KCl, 32.5 μM MgSO$_4$ 7H$_2$O, 54.5 μM CaCl$_2$ 2H$_2$O) at 25 °C. *A. proteus* was fed *Tetrahymena pyriformis* twice a week. *T. pyriformis* was cultured in 2% (w/v) Bacto Proteose Peptone medium for 3 days.

### Observation of cell migration

The migratory motion of *A. proteus* was monitored using dark-field microscope (Eclipse Ti, Nikon, Japan) with x 1 objective lens (CFI Plan UW 1X, Nikon, Japan) and CMOS camera (ORCA-Flash4.0, Hamamatsu Photonics, Japan). Red light (632 nm±11 nm) was used for illumination.

### Experimental procedure

After feeding *T. pyriformis*, we waited for 3 h 20 min. Then, the medium was replaced with KCM to wash out the remaining *T. pyriformis*. After 0, 1 and 3 days, *A. proteus* was transferred to a glass bottom dish. We started to record video with 1 fps after 1 h shading.

### Analysis of cell trajectories and cell shape

Trajectories and the shape of *A. proteus* were determined and analyzed using MATLAB software (see detail in (25)). Based on the edge detection of the cell, we extracted the shape of the cell from the dark field images. Then, the centroid $\vec{x}(t)$ of the cell was calculated. Before calculating velocity $v_1$ and $\dot{C}_n$, we took the moving filter of $\vec{x}(t)$ and $C_n(t)$ over 5 consecutive data points. Velocity and $\dot{C}_n(t)$ were calculated as $\vec{v}(t) = (\vec{x}(t+\tau) - \vec{x}(t-\tau))/2\tau$ and $\dot{C}_n(t) = (C_n(t+\tau) - C_n(t-\tau))/2\tau$ with $\tau$ = 30 s. Then we also averaged $C_n(t)$ as $(C_n(t+\tau) + C_n(t-\tau))/2$.

### Fitting of the experimental data by the models

$\beta_1$, $\beta_2$ and $\beta_c$ in Eq. (2) were estimated by minimizing the squared residual error $S = \sum |V - v_{fluc}|^2$ between $V$ obtained experimentally and the estimated velocity $v_{fluc}$. To estimate the coefficients of

Eq. (4), we numerically search the set of fitting parameters that locally minimize the squared residual error $S = \sum |V - v_1|^2$. Note that it is rather difficult to analytically solve Eq. (4), which is non-linear equation of $v_1$. Thus, we numerically calculated the steady state value of $\dot{v}_1 = -v_1 + v_{fluc} + v_{blls}$ by using $C_n$ and $\dot{C}_n$ for every time step. Then, we minimized the squared residual error by local random search of the coefficients.

**Akaike Information Criterion of the models**

AIC is calculated as
$$AIC = M \log(2\pi S/M) + M + 2(k + 1)$$
where $M$, $S$, and $k$ are the number of data points, squared residual error, and the number of fitting parameters, respectively (46). Generally, a smaller AIC is better than a larger one. In this work, $M$ is much larger than $k$ ($M \sim 10^3$, $k \sim 1$). Thus, squared residual error dominantly determines AIC.

**Acknowledgments**
This work was performed under the Cooperative Research Program of the Network Joint Research Center for Materials and Devices (H.E., Y.N., S.K., M.I.) and the Program for Fostering Researchers for the Next Generation conducted by the Consortium Office for the Fostering of Researchers in Future Generations, Hokkaido University (Y.N.). This work was supported by Japan Society for the Promotion of Science (JSPS) KAKENHI Grant numbers 19KK0180 (Y.N.), 21H05308 (Y.N.), 21K18326 (S.K.), 21K03855 (M.I.), 22H05678 (M.I.),  the JSPS Core-to-Core Program, an Advanced Research Networks (Y.N.), JPJSCCA20230002 (H.E., M.I.),  the Project of Junior Scientist Promotion at Hokkaido University (Y.N.), and JST, CREST Grant number JPMJCR22B1, Japan (S.K.).